\documentclass[%
 reprint,
 amsmath,amssymb,
 aps,
 pra,
]{revtex4-1}

\usepackage{graphicx}
\usepackage{dcolumn}
\usepackage{bm}
\usepackage{natbib}
\usepackage{siunitx}
\usepackage{textcomp}
\usepackage{color}
\usepackage{tikz}
\usetikzlibrary{shapes}
\usepackage{hyperref}

\begin{document}

\preprint{APS/123-QED}

\title{Distance scaling and polarization of  electric-field noise in a surface ion trap}

\author{Da An}
\thanks{These authors contributed equally to this work.}
\author{Clemens Matthiesen}
\thanks{These authors contributed equally to this work.}
\author{Erik Urban}
\author{Hartmut H\"{a}ffner}

\affiliation{%
 Department of Physics, University of California, Berkeley, CA 94720
}%

\date{\today}

\begin{abstract}
We probe electric-field noise in a surface ion trap for ion-surface distances $d$ between 50 and 300~\textmu{}m in the normal and planar directions.
We find the noise distance dependence to scale as $d^{-2.6}$ in our trap and a frequency dependence which is consistent with $1/f$ noise. Simulations of the electric-field noise specific to our trap geometry provide evidence that we are not limited by technical noise sources.
Our distance scaling data is consistent with a noise correlation length of about 100 \textmu m at the trap surface, and we discuss how patch potentials of this size would be modified by the electrode geometry.
\end{abstract}

\pacs{Valid PACS appear here}
\maketitle


\section{\label{sec:intro}Introduction}
Trapped atomic ions have emerged as a promising platform for the physical realization of quantum computation and simulation \cite{Leibfried2003review}.
However, a prominent challenge for the miniaturization of the traps necessary for scaling 
is the presence of electric-field noise. This noise causes decoherence of the ion motional modes and limits the fidelity of multi-ion qubit interactions \cite{Wineland1998-experimental-issues}.
The electric field noise originates from surfaces close to the trapped ion and is typically orders of magnitude larger than the Johnson noise expected from the thermal motion of the electrons in a  conductor of the same geometry \cite{Brownnutt2015}. While identifying the microscopic sources of this excess electric-field noise is an active topic of research, it has been found to scale strongly with distance to the trap surface \cite{Brownnutt2015, Turchette2000}. The precise scaling with distance provides information on the spatial extent of surface noise sources and is hence relevant to understanding the underlying mechanisms behind excess surface noise. This information may, in turn, allow the community to engineer better ion traps or inspire reliable surface treatments. Beyond ion traps, surface electric-field noise limits the performance of solid-state quantum sensors, such as nitrogen-vacancy centers in diamond \cite{Kim2015}, and hinders precision measurements, including gravitational probes with charged particles \cite{Camp1991MacroscopicConductors, Darling1992TheProblems} and Casimir-Polder force studies \cite{Sandoghdar1992DirectCavity, Harber2005MeasurementCondensate}. Thus, a better understanding of surface noise in ion traps may help advance research in these fields. 

Electric-field noise is quantified by its spectral density $S_E$ which we consider as a function of frequency $\omega$ and ion-surface distance $d$. The dependence on both variables is often described by a local power-law dependence, such that $S_E(d) \propto d^{-\beta(d)}$ for the distance scaling. To date, various possible sources of excess electric-field noise have been proposed and studied theoretically, resulting in a range of possible $\beta$ values from $0-8$, depending on both the noise source and the trap geometry \cite{Brownnutt2015}.

A small number of experiments have directly measured electric-field noise as a function of the ion-surface distance so far. The first two of these were performed in traps with non-planar geometries, using needle-shaped tungsten electrodes in one case \cite{Deslauriers2006a} and a gold-plated `stylus trap' design in the other \cite{Arrington2013Micro-fabricatedTrap}. These experiments found electric-field noise scalings of $\beta = 3.5$ \cite{Deslauriers2006a} and $\beta = 3.1$ \cite{Hite2017MeasurementsProximity}, respectively. More recently, two further experiments employed planar surface traps for distance scaling measurements of noise parallel to the trap surface. The first study used a standard five-wire trap with gold electrodes and non-standard radio-frequency (RF) voltages and obtained a distance-scaling exponent of $\beta = 3.8$ \cite{Boldin2018}. The magnitude of electric-field noise in this experiment was about an order of magnitude higher than the best published results for untreated surface traps.
The second study found $\beta = 4.0$ for a multizone niobium surface trap \cite{Sedlacek2018}. Here, the ion trap contained several trapping zones spaced by on the order of 1~mm, where the electrodes at each zone were scaled to trap ions at different distances from the surface. The magnitude of electric-field noise was comparable to very good untreated surface traps.

Collectively, these previous measurements are consistent with a noise mechanism of microscopic nature which appears to be common to ion traps of different materials and geometries.

In our experiment, we investigate the electric-field noise as a function of ion-surface distance for a single $^{40}$Ca$^+$ ion trapped in a unique surface Paul trap with a simple planar geometry. Our room-temperature four-RF-electrode trap enables tuning of the ion-surface distance at a fixed planar position using DC voltages  \cite{An2018}, and we find that the magnitude of noise for this trap is comparable to the best untreated traps. We measure electric-field noise in both the normal and planar directions with respect to the surface and extract a distance scaling exponent of $\beta \approx 2.6$ for ion-surface distances in the range of $50-300$ ~\textmu m, significantly departing from the previous measurements discussed above. We provide evidence that the measured field noise is not limited by technical noise (defined to be noise whose origin is external to the properties of the trap surface, for example, from power supplies or electromagnetic pickup), discuss spatial correlations of noise 
implied by our data, and place the results in a larger context with respect to previous measurements.

In section \ref{sec:methods} we describe the experimental setup and measurement methods. Section \ref{sec:results} contains the measurement results for distance and frequency scalings. Section \ref{sec:technoise} provides a discussion around technical noise, and in section \ref{sec:theory} we discuss how spatial correlations of the field noise on the trap surface may explain the observed distance dependency. We summarize and conclude in section \ref{sec:conclusion}.

\begin{figure}[t!]
    \centering
    \includegraphics[scale=.25]{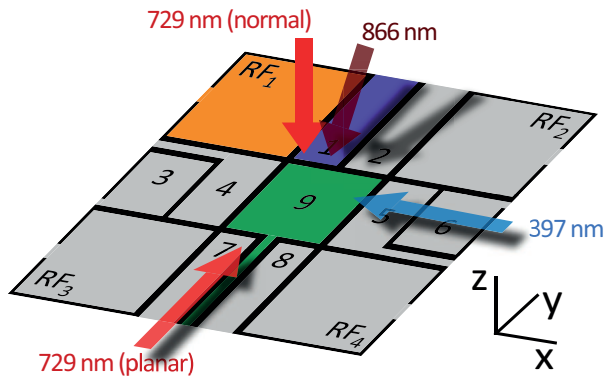}
    \caption{Isometric view of the trapping region of the four-RF-electrode trap. Static voltages are applied to numerically labeled DC electrodes, and RF electrodes are labeled $\mathrm{RF_{1-4}}$. $\mathrm{RF_{1,4}}$ are driven out-of-phase relative to $\mathrm{RF_{2,3}}$. Technical noise simulations are performed on colored electrodes \#RF$_1$, \#1, \#9 (see Sec.~\ref{sec:technoise}). The orientation of the qubit addressing laser beam (729~nm) and Doppler cooling beams (397~nm and 866~nm) are indicated by the arrows and their labels.}
    \label{fig:trap}
\end{figure}

\section{\label{sec:methods}Experimental Setup and Methods}
\subsection{\label{sec:trapdesign}Trap Design}
Our four-RF-electrode trap is fabricated by laser-etching fused silica (Translume, Ann Arbor, Michigan), and double-layering the surface with titanium (15~nm), aluminum (500~nm), and copper (30~nm) using electron-beam physical vapor deposition. The trap is operated with an out-of-phase RF drive such that the two diagonal pairs of RF electrodes ($\mathrm{RF_{1,4}}$ and $\mathrm{RF_{2,3}}$) are driven with the same amplitude, but opposite phase.
This generates a radio-frequency-null along the axis normal to the trap surface \cite{An2018}. Thus, the trapping potential in this direction is fully controlled with DC electrodes (1 to 9), allowing for continuous variation of the ion-surface distance without introducing excess micromotion. RF fields provide confinement in the $xy$ plane parallel to the trap. In analogy to a typical linear surface trap, the motional mode normal to the surface then corresponds to the axial mode, while the planar modes correspond to the radial modes.

In addition to tuning the ion-surface distance, the DC fields are also used to generate quadrupole potentials breaking the rotational symmetry of the RF potential around the $z$ axis and tilting principal axes of the harmonically trapped ion (see Fig.~\ref{fig:trap} for coordinate system). We adjust the DC voltages such that the principle axis of the total potential have a tilt of $5^{\circ}$ with respect both the $z$ and $y$-axes. This ensures that all three motional modes have some projection onto the $397$~nm Doppler cooling beam propagating along $\hat{x}$. To provide additional cooling for the near-normal mode, the repumping $866$~nm beam is sent in at a near-normal angle with respect to the trap surface ($z$ axis) and detuned red with respect to resonance.

The beam at 729-nm wavelength addressing the long-lived quadrupole qubit transition of $^{40}$Ca$^+$ can be switched between two orientations to measure the ion's response to electric-field noise in two directions. The `729~nm (normal)' orientation in Fig.~\ref{fig:trap} is used for measuring noise normal to the trap surface (along the $\hat{z}$-direction), while the `729~nm (planar)' orientation allows noise measurements in the $\hat{y}$-direction of the trap plane.

The voltages applied to the DC electrodes (labeled 1-9 in Fig. \ref{fig:trap}) are low-pass filtered with a cut-off frequency below $300$~Hz before going into the vacuum chamber, and filtered again inside the chamber with two grounding capacitors of 47~nF and 10~nF. The RF signal is filtered with a 10~MHz high-pass before going to an inductively coupled toroidal resonator connected to the trap electrodes \cite{An2018}.
Importantly, we find that careful grounding of the vacuum chamber, the RF voltage source, and the optical table on which the vacuum chamber is mounted is essential for low-noise measurements. Without these ground connections electric-field noise in the direction normal to the trap surface typically increases by 1-3 orders of magnitude, while noise in the planar direction increases by less than a factor 10.

\subsection{\label{sec:hrmethods}Measurement Methods}
Fluctuating electric fields at the ion location induce heating of the motional modes, which is quantified by the heating rate $\dot{\bar{n}}$ \cite{Brownnutt2015}:
\begin{equation}\label{eq:nbardot_vs_SE}
\dot{\bar{n}} = \frac{e^2}{4m\hbar\omega} S_E(\omega, d)\;,
\end{equation}
where $\bar{n}$ is the average motional mode occupation, $e$ is the ion charge, $m$ is the ion mass, $\hbar$ is the reduced Planck constant, and $\omega$ is the angular secular frequency. Thus, the spectral density $S_E$ of the electric-field noise may be inferred through measurement of the ion heating rate.

As the strength of electric-field noise changes by two orders of magnitude over the ion-surface distance explored here, we employ two methods for measuring heating rates. When the ion is cooled close to the motional ground state, that is $\bar{n}\leq 1$ (possible if $\dot{\bar{n}}\lesssim 1000~\mathrm{s}^{-1}$ for our system), the amplitude of the phonon-subtracting transition (red sideband) is significantly less than that of phonon-adding  one (blue sideband). Then, the mean phonon number $\bar{n}$ can reliably be found by comparing the excitation amplitude of red and blue first-order motional sidebands~\cite{Monroe1995a}. When cooling into the ground state is difficult ($\dot{\bar{n}} \gtrsim 1000~\mathrm{s}^{-1}$), the sideband method is no longer sensitive to $\bar{n}$. In this regime, a better measure of the heating rate is via the decay of carrier Rabi oscillations~\cite{Rowe2002}, which can be fitted to extract the thermal motional distribution parameterized by $\bar{n}$ and thereby the heating rate $\dot{\bar{n}}$.

\newcommand{\markersquare}{\raisebox{0.5pt}{\tikz{\node[draw,scale=0.4,regular polygon, regular polygon sides=4,fill=black](){};}}}
\newcommand{\markerdiamondfilled}{\raisebox{0pt}{\tikz{\node[draw=gray!40!white,scale=0.4,diamond,fill=gray!40!white](){};}}}
\newcommand{\markerdiamondunfilled}{\raisebox{0pt}{\tikz{\node[draw,scale=0.4,diamond,fill=none,thick](){};}}}

\begin{figure}[t!]
    \centering
    \includegraphics[scale=1]{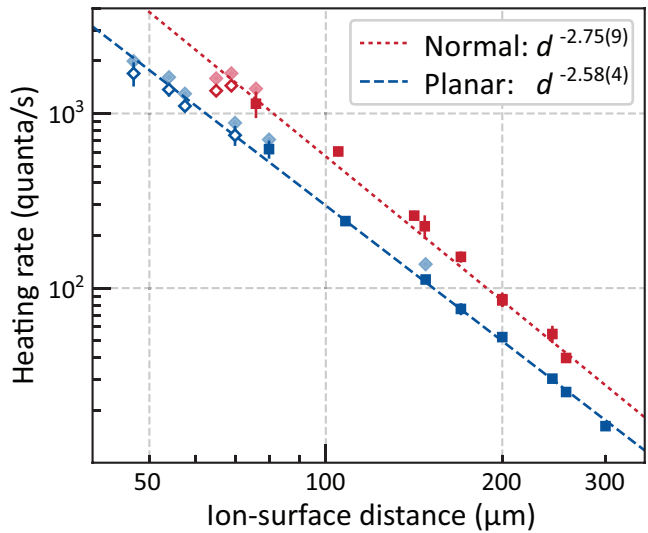}
    \caption{Planar (blue) and normal (red) heating rates as a function of ion-surface distance for a fixed secular frequency of $\omega_t = 2\pi \times 1$~MHz. Data are taken with two measurement methods: (\protect\markersquare) sideband method, (\protect\markerdiamondfilled) Rabi method. (\protect\markerdiamondunfilled) show data from the Rabi method scaled to match results from the sideband method (see comparison in Sec. \ref{sec:results}). Power-law fits for both motional modes take into account the data taken with the sideband-asymmetry (\protect\markersquare) method and the scaled Rabi method (\protect\markerdiamondunfilled).}
    \label{fig:hrscaling}
\end{figure}

\begin{figure}[t]
    \centering
    \includegraphics[scale=1]{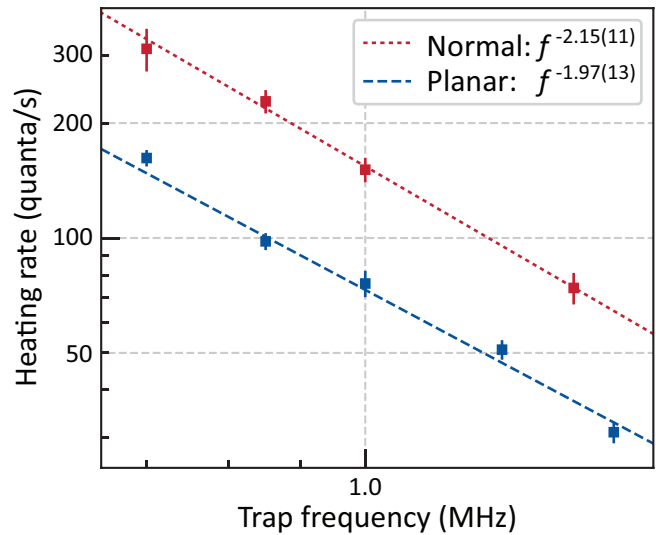}
    \caption{(b) Planar (blue) and normal (red)  heating rates as a function of secular frequency for a fixed ion-surface separation of $d = 170$~{\textmu}m. Power-law fits for both modes are shown as dashed and dotted lines.}
    \label{fig:freqscaling}
\end{figure}

\section{\label{sec:results}Results}
\subsection{\label{sec:distancescaling}Distance Scaling}
The scaling of electric-field noise with $d$ is determined from measurements of the heating rate at a fixed secular frequency of $f = 2\pi\times1.00$~MHz.
The two planar motional modes are closely aligned with the $x$ and $y$-axes (see Sec. \ref{sec:methods}) and we measure the heating rate of the $\hat{y}$ planar mode. Using DC-voltages, the $\hat{x}$ planar mode is set higher in frequency by typically $ 2\pi\times50$~kHz than the $\hat{y}$ mode frequency.
When measuring the planar mode at $2\pi\times1$~MHz, the normal $\hat{z}$ mode is set to about $2\pi\times0.75$~MHz, while for measurements of the normal mode at $2\pi\times1$~MHz, the planar mode frequencies are tuned to about $2\pi\times1.25$~MHz. The range of ion-electrode distances accessible with the trap is limited by the high RF voltages needed when trapping close to and far away from the surface \cite{An2018}. 
We also find that for measurements of the normal mode, care must be taken to reduce back-reflections of the normal 729~nm beam from the trap surface, which cause significant intensity fluctuations at the ion position, leading to unreliable measurements especially for small distances from the trap surface.

The heating rates for the planar (blue symbols) and normal (red symbols) mode are plotted in Fig.~\ref{fig:hrscaling}. We distinguish between the two measurement methods mentioned earlier: filled squares (\protect\markersquare) correspond to measurements taken with the sideband method. Below $80$~{\textmu}m ion-surface distance, the carrier Rabi-oscillation method is used instead, and the raw heating rate data are marked with filled diamonds (\protect\markerdiamondfilled).

To validate the compatibility of the two methods, we compare both measurement techniques in the intermediate regime of $\dot{\bar{n}} \sim 1000~ \mathrm{s}^{-1}$ and also for $\dot{\bar{n}} \sim 100~ \mathrm{s}^{-1}$. We find that the Rabi method gives heating rates that are systematically about $17\%$ higher than for the sideband-asymmetry method. This difference is consistent with findings from other experiments \cite{Shu2014HeatingTrap, Sedlacek2018} and may be linked to the non-zero projection of other motional modes on the measurement laser or laser intesnity noise.
To simplify comparison of the electric-field noise across the full range of ion-surface distances, we scale the data from the Rabi-oscillation method by a constant factor of $0.85$. The resulting adjusted values are marked in Fig.~\ref{fig:hrscaling} as open diamonds (\protect\markerdiamondunfilled).

Power-law fits to the filled squares (\protect\markersquare) and open diamonds (\protect\markerdiamondunfilled) are shown as dashed lines in Fig. \ref{fig:hrscaling}. We observe the scaling  $\dot{\bar{n}} \sim d^{-2.6}$ for both modes, with normal heating rates consistently about a factor two higher than planar heating rates. While our data are described fairly well by a power-law behaviour, there may be some systematic deviations to the power-law behavior at low ion-electrode distances.

\begin{figure*}[t!]
    \centering
    \includegraphics[scale=1]{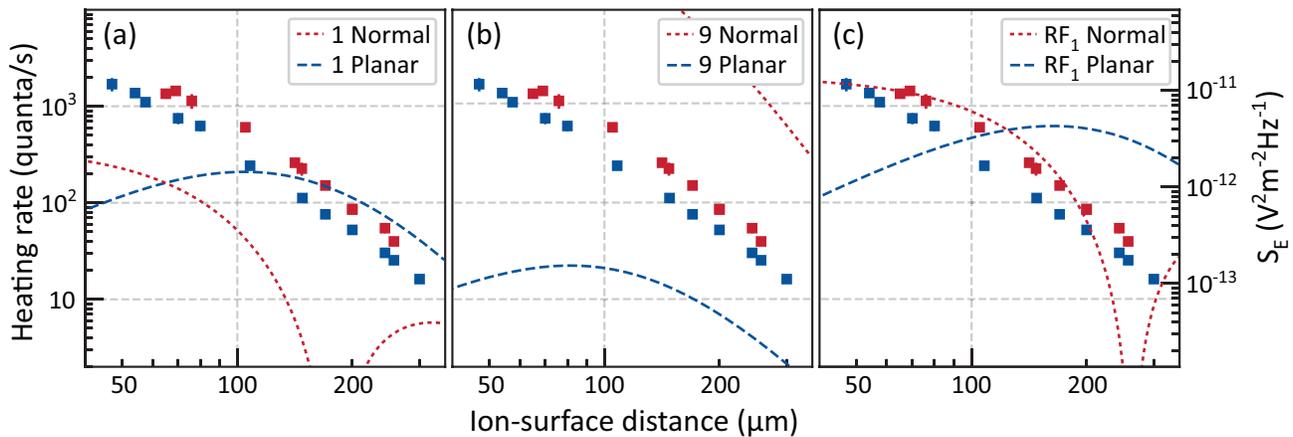}
    \caption{Simulated $S_E$ distance scaling from $3$-{\textmu}V amplitude noise on electrodes 1 (a), 9 (b), and $\mathrm{RF_1}$ (c). The planar contributions are shown as blue dashed curves and the normal contributions are shown as dotted red curves. Scaled data measurements are overlaid for reference.}
    \label{fig:technoisesims}
\end{figure*}

\subsection{\label{sec:freqscaling}Frequency Scaling}
In addition to the distance scaling, we measure the scaling of noise with frequency, which is another key parameter to characterize the electric-field noise. In Fig.~\ref{fig:freqscaling} we show the noise frequency dependence of both normal and planar modes at a fixed ion-surface distance of $170$~{\textmu}m, where all measurements are performed with the sideband-asymmetry method. From a power-law fit of the data and Eq.~\eqref{eq:nbardot_vs_SE}, we find that electric-field noise spectral density, $S_E(2\pi\times f, d)$, scales as $f^{-0.97(13)}$ and $f^{-1.15(11)}$ for the planar and normal modes, respectively. This frequency dependence is consistent with $1/f$-noise that is ubiquitous in solid-state experiments \cite{Paladino20141/Information} and a number of ion trapping experiments \cite{Brownnutt2015}. We observe again that the normal heating rates are about a factor of two higher than the planar heating rates. Additionally, we measure the frequency scaling of the planar electric-field noise at 70~{\textmu}m ion-surface distance and find $S_E(2\pi\times f, d) \propto f^{-1.2(3)}$. The $1/f$-like noise scaling for different ion-surface distances and electric-field projections in our system suggests that our measurements are not described by proposed models for noise from adatom dipole fluctuations \cite{Safavi-Naini2011} or adatom diffusion \cite{Kim2017Electric-fieldExperiments}, or by broadband technical noise \cite{Brownnutt2015} since the expected frequency scaling in these cases is not $1/f$.

Having presented our measurement results, we note that the distance scaling for this ion trap differs strongly from the $d^{-4}$ dependence that for planar trap geometries is expected  from microscopic noise sources and has been observed in recent experiments \cite{Boldin2018,Sedlacek2018}. A scaling exponent of $\beta \approx 2$ is commonly associated with technical noise; however, this scaling only appears if the trap size is scaled with the ion-surface distance. This is not the case in our system. Thus, understanding how technical noise manifests in this trap and determining to what degree it affects our measurements are critical issues that we address in the following.


\section{\label{sec:technoise}Technical Noise Checks}
Two types of electric-field noise sources can affect the motional ion heating: surface noise and technical noise. Surface noise encompasses the mechanisms originating from the ion trap surface itself, that is, physical processes generating noise at the trap surface. Technical noise is defined as noise whose origin is external to the properties of the trap surface, and thus can be mitigated by proper filtering and shielding. Technical noise may be caused by noise from voltages sources powering the ion trap or electromagnetic radiation in the environment. This type of noise manifests itself as correlated over each trap electrode, and thus depends specifically on the trap electrode geometry and the ion's position relative to the electrodes. As we are interested in the fundamental physical noise sources that could limit the performance of trapped ion qubits, it is important that our electric-field noise measurements are dominated by surface noise. Therefore we perform several checks to verify that our heating rates are not limited by technical noise.

\subsection{\label{sec:electrodenoisesims} Electrode Noise Simulations}
The assumption that technical noise is correlated over the area of each trap electrode is valid in our system because the average dimension of our electrodes is $100$~{\textmu}m while the measured noise is at a frequency of $f = 2\pi\times1$~MHz, corresponding to a wavelength of about $300$~m.
The spectral density of technical electric-field noise is $S_E \propto \sum_{i}E_{i}^2$, where $E_{i}$ is the electric-field due to electrode $i$ at the ion location.

For each trap electrode, we simulate $S_{E,i}$ as a function of the ion-surface distance under the assumption that technical noise dominates at the ion position. $E_{i}$ is analytically approximated by a method which assumes an infinite ground plane surrounding electrode $i$ \cite{Wesenberg2008}. From the geometry of our trap (cf. Fig.~\ref{fig:trap}), we see that only electrodes 1, 2, 7, 8, 9, and $\mathrm{RF_{1-4}}$ generate electric fields with projection onto both the $y$ and $z$ axes, which correspond to the directions of the measured ion motional modes. Also, from the symmetry of our trap, we expect electrodes 1, 2, 7, and 8 to generate similarly polarized electric-field noise in the $yz$ plane, as is the case for all four RF electrodes. So, we will discuss here the distance scaling for the relevant cases of electrodes 1, 9, and $\mathrm{RF_{1}}$. Fig.~\ref{fig:technoisesims}(a)-(c) visualizes the simulated electric-field noise in planar and normal directions for $3$-{\textmu}V noise amplitude on each electrode. Comparing the simulations with the overlaid scaled data, we find none of the three cases match the data. Similarities are only partial and do not apply simultaneously to both projections of noise. In general, for correlated noise from trap electrodes, the normal-planar polarization is not constant as a function of the ion-electrode distances, in contrast to our data. Thus, we conclude that no single electrode can generate noise consistent with our measured electric-field noise data. 

\begin{figure}[t!]
    \centering
    \includegraphics[scale=1]{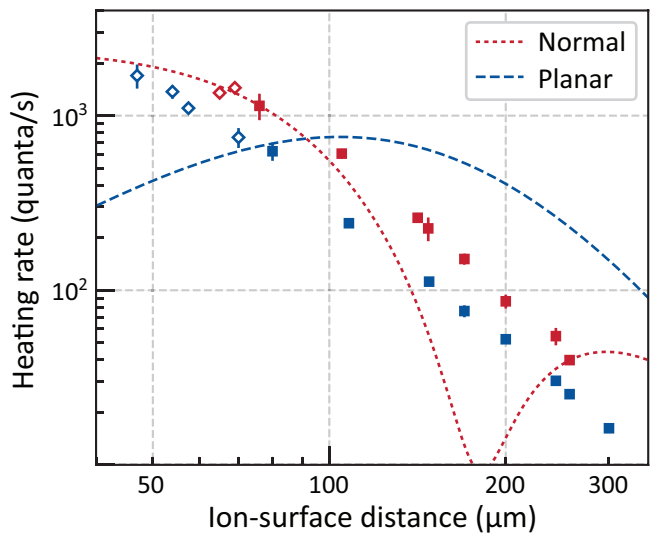}
    \caption{Simultaneous fit of the measured distance dependence for planar and normal motional modes assuming motional heating is caused by technical noise from all electrodes. The voltages on different electrodes are assumed to fluctuate independently from each other such that the fit is parameterized by the amplitudes of noise on each electrode.}
    \label{fig:technoise_conspiracy}
\end{figure}

Beyond noise from individual electrodes, one could imagine technical noise contributions with various amplitudes from all electrodes to play a role, conspiring to give rise to the particular distance dependence we measure.
To check the possibility of such a scenario we fit the measurements with a linear combination of contributions from all electrodes, assuming noise from each electrode is uncorrelated with the others. Even for this contrived scenario, see fit in Fig.~\ref{fig:technoise_conspiracy}, we do not find any agreement with our data.

Given the strong disagreement between the simulated noise from electrodes and our experimental results, we conclude that our measurements are not limited by technical noise correlated across trap electrodes. These arguments also hold for Johnson noise originating from the voltage supply and filter electronics, since we expect Johnson noise of this type to be correlated over the area of an electrode as well.


\section{\label{sec:theory}Spatial Surface Noise Correlations}

In light of the strong evidence for uncorrelated microscopic noise sources in ion traps given by the observed $d^{-4}$ scaling \cite{Boldin2018, Sedlacek2018, Sedlacek2018a}, it is surprising that our results ($\sim d^{-2.6}$) deviate so clearly from this scaling. We have shown in the last section that measurements in our trap are very likely not limited by technical noise, leading us to conclude that we must be observing surface noise.
There are immediate questions as to how such a scaling can arise, what the noise sources are, and why this trap behaves differently compared to others. In the following we concentrate on the first of these questions.

A general model for surface noise with finite correlations considers metallic surfaces to be covered with patches of varying potential \cite{Turchette2000}. Physically, patches may arise from work function differences across crystal grains, locally varying strain, surface roughness, or adsorption of atoms and more complex compounds on the surface \cite{Herring1949ThermionicEmission}. Fluctuations in the patch potentials (amplitude or size) lead to electric-field noise then. In the `patch-potential' picture, the behavior of electric-field noise is quantified by a spatial length scale, $\zeta$, over which the surface fluctuations are correlated. While the model exists independently of the physical origins of the patches, a certain correlation length may indicate make the explanation by some noise sources more likely than by others.

In the limit of spatial correlations that are very small compared to the ion-surface distance, $\zeta \ll d$, the electric-field noise from different locations on the surface stems from independent microscopic noisy patches and adds in quadrature at the ion position. Then, for a planar trap geometry, the electric-field magnitude scales as $d^{-4}$ with distance. Further, there is a polarization to the electric-field noise such that the projection on the normal direction is twice that for the planar directions \cite{Low2011,Schindler2015}.
In the opposite limit of very large spatial correlations, $\zeta \gg d$, the fields only vary weakly with distance from the surface. Depending on the specific arrangement of patches or the form of the spatial correlation function, the noise may, for instance, be independent of distance or scale as $d^{-1}$ and the polarization of normal and planar directions can be very large. At intermediate distances, the scaling coefficient varies smoothly from one limit to the other.

Measurements of the electric-field noise distance scaling then probe the spatial extent of noise correlations on the surface. With regard to our data, the measured scaling exponent of $\beta = 2.6$ being smaller than $\beta = 4$ may be taken as an indication of macroscopic correlation lengths, as compared to the ion-surface distance. To further understand the role of the correlation length $\zeta$ in describing our data, we will review a general framework for modeling correlated patch-potentials and also give an example for a specific physical realization of patch potentials in our surface trap.

\begin{figure}[t!]
    \centering
    \includegraphics[scale=1]{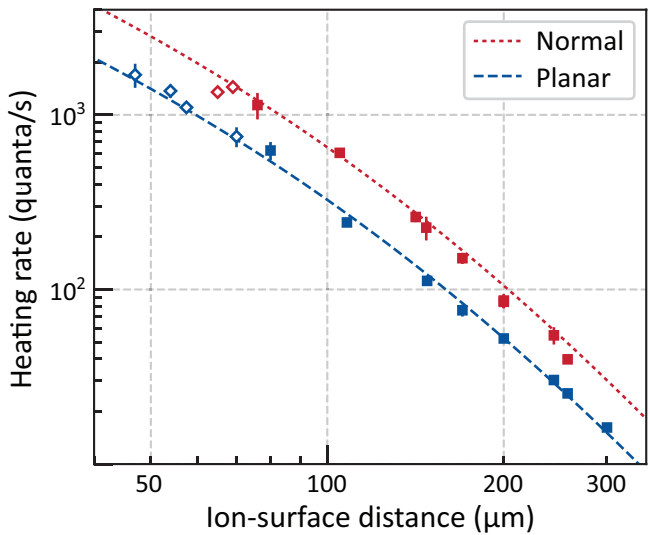}
    \caption{Heating rate as a function of distance with a fit  based on exponential spatial noise correlations. Fitting both data sets to Eq. \eqref{eq:S_E_exponential} gives a  characteristic length $\zeta = 106$~{\textmu}m.}
    \label{fig:dscaling_corr_patch}
\end{figure}

\begin{figure*}[t!]
    \centering
    \includegraphics[scale=1]{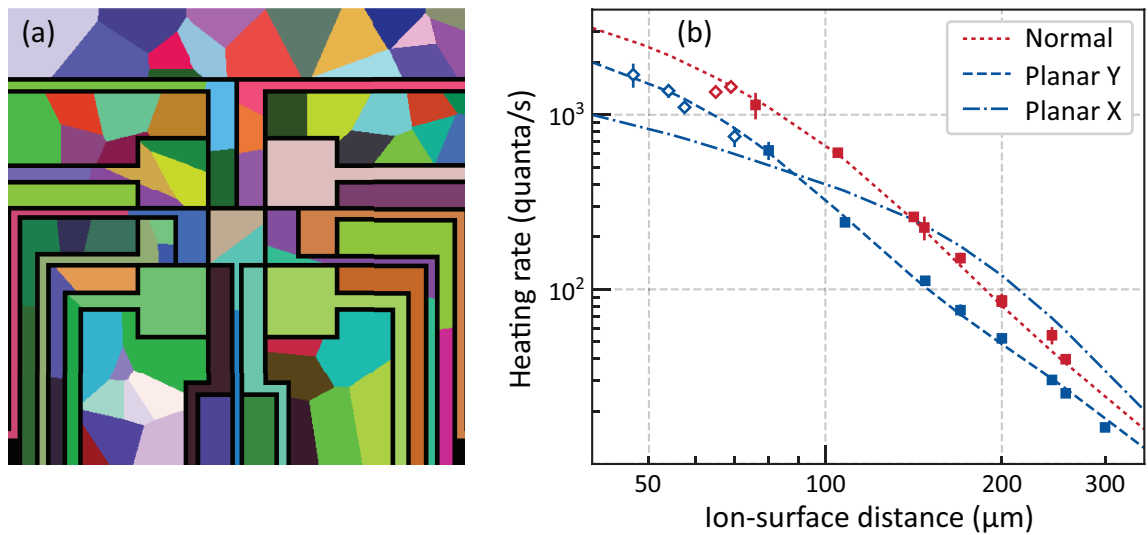}
    \caption{(a) Example of randomly generated patch configuration based on Poisson-Voronoi tesselation with correlation length $\zeta = 140$~{\textmu}m. (b) Distance scaling data fit with patch configuration in (a) and variable voltage amplitude on the patches. The ratio between the largest and smallest voltage is four. Note the noise anisotropy in the planar directions.}
    \label{fig:dscaling_fixed_patch}
\end{figure*}

\subsection{\label{sec:correlatedpatchpotential}Analytic Model}
Starting from the characteristic length, $\zeta$, over which the noisy potential patches are correlated, we can explicitly calculate the electric-field noise spectral density vector following Refs.~\cite{Dubessy2009, Low2011, Dehmelt1968}.
We assume an exponential spatial autocorrelation function,
\begin{equation} \label{eq:autocorrelation}
C_\zeta = e^{-\sqrt{x^2+y^2}/\zeta},
\end{equation}
which arises naturally from random variations of a variable in two or three dimensions, for instance in Poisson-Voronoi tessellations \cite{Debye1957ScatteringApplication, Man2006OnMaterials}.
Then, the expected planar noise spectral density $S^{p}_E$ is given by
\begin{equation} \label{eq:S_E_exponential}
S^{p}_E(\omega_t,d) = 2 \frac{N \zeta^2}{d} S_V(\omega_t) \int_{0}^{\infty} dk \frac{k^3 e^{-2k}}{(d^2 + \zeta ^2 k^2)^{3/2}}\; ,
\end{equation}
and the spectral density of noise in the normal direction is higher by a factor of two.

With this analytic model, we find that for $\zeta \ll d$, we recover $d^{-4}$ scaling for both the planar and normal heating rates, as expected from similar microscopic noise models. As the ratio between $d$ and $\zeta$ inverts, there is a smooth transition in the exponent of the power law towards the limiting case of $\zeta \gg d$, where the distance scaling approaches a $d^{-1}$ behavior.

We use Eq. \eqref{eq:S_E_exponential} to simultaneously fit both the measured planar and normal distance scalings (see Fig.~\ref{fig:dscaling_corr_patch}), and extract a correlation length of $\zeta = 106$~{\textmu}m. This provides a good fit to the data, that is comparable to the power-law fit with $\beta = 2.6$ from Fig.~\ref{fig:hrscaling}.

While we can describes our data with a single parameter, the correlation length, the same autocorrelation function can arise from many distinct patch configurations. The curves in Fig. \ref{fig:dscaling_corr_patch} effectively show the distance scaling for an average over all patch configurations with the same correlation length. Reproducing this exact scaling with well-defined patches at the trap surface would require either a superposition of overlapping patches or fast switching of patch configurations.

Neither scenario is likely to manifest on a simple metallic surface, but the trap used here has a considerably more complex structure. The combination of features like an oxide layer \cite{Kumph2016} on the aluminium-copper electrodes, adsorbates on the surface, and the influence of surface roughness \cite{Lin2016EffectsIons}, may allow for more complex arrangements of charges and fluctuating dipoles. The presence of an insulating layer (the oxide), for example, could separate patch potentials both above and below it, creating a structure of overlapping patches.
One might also imagine dynamic patches that shift, rearrange or reassemble on timescales much faster than the experiment (of order one second) \cite{VanGastel2001NothingDiffusion}. Measuring the surface potentials of this trap directly with a technique like Kelvin probe or scanning tunneling microscopy may provide additional insight on the noise origin.

The general correlation length model used here also assumes an infinite planar metallic surface, which does not quite translate to our surface trap that is composed of many planar electrodes separated by $20$-{\textmu}m wide trenches. Such gaps between conducting surfaces should act as natural boundaries over which noise correlations cannot be established.

\subsection{\label{sec:fixedpatchpotential}Fixed Patch Potentials}
For a more concrete realization of a noise correlation length, we consider non-overlapping patches of fixed size and position while imposing the constraint that correlations cannot form across electrode boundaries. We generate many random patch configurations based on two-dimensional Poisson-Voronoi diagrams. An example configuration with $\zeta = 140$~{\textmu}m is shown in Fig.~\ref{fig:dscaling_fixed_patch}(a) where the correlation length, $\zeta$, is extracted from an exponential fit to the calculated spatial autocorrelation function.

Each patch is assumed to generate noise independently of other patches and the noise amplitude on each patch is a parameter for fitting the patch potential distance scaling to our data. The fit result for the example patch configuration in Fig.~\ref{fig:dscaling_fixed_patch}(a) is shown in Fig.~\ref{fig:dscaling_fixed_patch}(b). A consequence of moving from the analytic correlation length model (previous subsection, Fig. \ref{fig:dscaling_corr_patch}) to a defined tiling of patches is a general anisotropy of noise in different planar directions, as exemplified by the noise spectral densities `Planar X' and `Planar Y' in Fig. \ref{fig:dscaling_fixed_patch}(b). Here the simulated distance scaling matches the data well, but since we only compare to noise in the $y$-direction and normal to the surface, this is not a unique solution for the patch arrangement. Many patch configurations with similar average patch sizes reproduce fit results similar to the example given in Fig.~\ref{fig:dscaling_fixed_patch}. Among these well fitting configurations, we find that the center electrode, DC9, must contain at least two patches, else the distance scaling behaves similarly to technical noise case shown in Fig.~\ref{fig:technoisesims}(b).

Comparing the fixed patch configuration to the general correlation length picture, we observe that the constraints posed by the electrode geometry lead to slightly longer correlation lengths, but still $\zeta$ is about 100 \textmu{}m. The key difference, as explained earlier, lies in the expected polarization of electric-field noise. The general model in Sec.~\ref{sec:correlatedpatchpotential} is an effective averaging over many random configurations, without preferential patch orientations. Thus, the electric-field noise is isotropic in all planar directions. In contrast, the fixed patch configuration generally features different electric-field noise in the planar directions.

We measured the planar electric-field noise in both the $y$ and $x+y$-directions at an ion-surface distance of 170~\textmu{}m and found them to be the same within the experimental uncertainty, ruling out some patch configurations that otherwise fit the observed distance scaling in the $y$-direction alone.
Further work on measuring the electric-field noise in three dimensions would be needed in order to achieve a better understanding of the nature of noise correlations in this trap.


\section{\label{sec:conclusion}Summary and Conclusion}

\subsection{\label{sec:comparison} Comparison of Heating Rates}
\begin{figure}[t!]
    \centering
    \includegraphics[scale=.97]{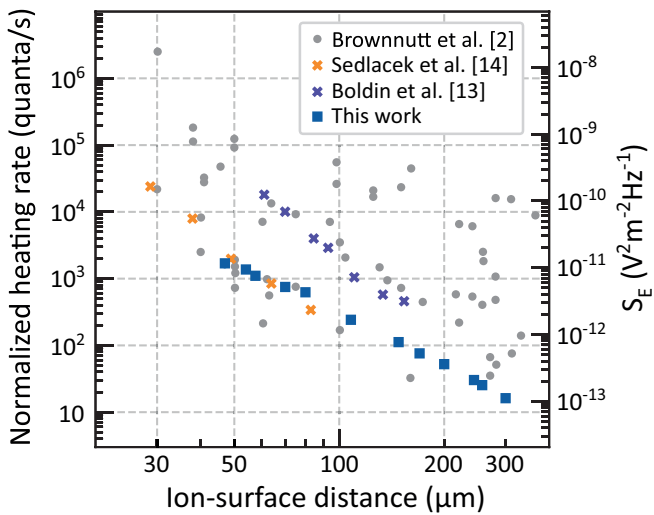}
    \caption{Comparison of the heating rates in this paper with those of other ion traps. The filled gray circles reproduce the data compiled in Ref.~\cite{Brownnutt2015} for single-ion motional heating, with the exception of surface-treated traps. Blue (dark) and yellow (light) crosses show the results of distance scaling measurements in the same trap from Ref.~\cite{Boldin2018} and Ref.~\cite{Sedlacek2018}, respectively.}
    \label{fig:comp}
\end{figure}

Finally we look at the magnitude of electric-field noise in our trap compared to other measurements in the community. Data collected in the review of Brownnutt et \textit{al.} \cite{Brownnutt2015} form the basis for the comparison, see Fig. \ref{fig:comp}, where we also included the recent data from Ref. \cite{Boldin2018} and Ref. \cite{Sedlacek2018} (highlighted as colored crosses), and removed results for surface-treated traps. Following the convention in Ref.~\cite{Brownnutt2015}, the heating rates are scaled on one axis to relate to the case of $2 \pi \times 1$~MHz secular frequency for a $^{40}$Ca$^+$ ion, and the other axis shows the electric field spectral noise density.

In comparison to all data on the plot, our heating rates are on the low end across the measured ion-surface distances. Looking specifically at the measurements for distance scalings performed in the same trap, our results are about a factor 5-20 lower than the ones reported by Boldin et \textit{al.} \cite{Boldin2018}, and of similar magnitude to measurements of Sedlacek et \textit{al.} \cite{Sedlacek2018}.

The similarities in the absolute noise magnitudes between the traps is also interesting, since one would generally expect that, given some density of microscopic noise sources, longer correlation lengths should increase the noise magnitude in our device.

Regardless of the physical origin of noise in the different traps, it is worth noting that with regards to device miniaturization, the distance scaling we observe ($d^{-2.6}$) scales favourably compared to the $d^{-4}$-dependence observed in other traps.
This observation provides us with some motivation to understand the origin of noise in this device and 
work towards further miniaturization of ion traps.

\subsection{Concluding remarks and summary}
To summarize, we have presented measurements of the distance scaling of electric-field noise in a surface ion trap, together with measurements of the frequency dependence. In contrast to previous results, the noise distance scaling for our trap is described by a $d^{-2.6}$ power-law behaviour. The data cannot be explained by noise from independent microscopic sources at the surface; we require the addition of a macroscopic length scale for the noise that extends to about 100~\textmu m.
The presence of a macroscopic length scale indicates a non-trivial surface structure and/or correlated dynamics of noise sources taking place at the trap electrode.
Our results add to the growing body of experimental results on electric-field noise in ion traps, and specifically show that the scaling of electric-field noise with distance is not universal in surface traps of similar size.

\pagebreak
\bibliography{references}

\end{document}